\documentclass[twocolumn,showpacs,pre,english,showkeys,aps,unsortedaddress,superscriptaddress,nofootinbib]{revtex4-2}

\usepackage{amsmath}
\usepackage{amssymb}
\usepackage[colorlinks]{hyperref}
\usepackage{epsfig}
\usepackage{graphicx}
\usepackage[T1]{fontenc}
\usepackage[utf8]{inputenc}
\usepackage[usenames,dvipsnames]{color}
\usepackage[normalem]{ulem}
\usepackage{tikz}
\usetikzlibrary{positioning}

\usepackage{multirow} 
\usepackage{enumitem}

\graphicspath{ {./figures/} }

\begin{document}

\title{Cluster percolation and dynamical scaling in the Baxter--Wu model}

\author{Alexandros Vasilopoulos}
\email{alex.vasilopoulos@essex.ac.uk}
\affiliation{School of Mathematics, Statistics and Actuarial Science, University of Essex, Colchester CO4 3SQ, United Kingdom}

\author{Michail Akritidis}
\email{michalis.akritidis@u-bourgogne.fr}
\affiliation{Institut de Math\'{e}matiques de Bourgogne, Universit\'{e} de Bourgogne Europe, 21078 Dijon, France}

\author{Nikolaos G. Fytas}
\email{nikolaos.fytas@essex.ac.uk}
\affiliation{School of Mathematics, Statistics and Actuarial Science, University of Essex, Colchester CO4 3SQ, United Kingdom}

\author{Martin Weigel}
\email{martin.weigel@physik.tu-chemnitz.de}
\affiliation{Institut für Physik, Technische Universität Chemnitz, 09107 Chemnitz, Germany}
\affiliation{Physics Department, Emory University, Atlanta, GA, U.S.A.}

\begin{abstract}
We investigate the percolation behavior of Fortuin-Kasteleyn–type clusters in the
spin-$1/2$ Baxter--Wu model with three-spin interactions on a triangular
lattice. The considered clusters are constructed by randomly freezing one of the
three sublattices, resulting in effective pairwise interactions among the remaining
spins. Using Monte Carlo simulations combined with a finite-size scaling analysis,
we determine the percolation temperature of these stochastic clusters and show that
it coincides with the exact thermal critical point of the model. The critical
exponents derived from cluster observables are consistent with those of the
underlying thermal phase transition. Finally, we analyze the dynamical scaling of
the multi-cluster and single-cluster algorithms resulting from the cluster
construction, highlighting their efficiency and scaling behavior with system size.
\end{abstract}

\date{\today}

\maketitle

\section{Introduction}
\label{sec:Introduction}

Cluster algorithms are a powerful tool for studying condensed-matter systems,
particularly in the vicinity of continuous phase
transitions~\cite{landau_book}. Owing to the nonlocal nature of their update moves,
they can substantially reduce---and in some cases practically eliminate---critical
slowing down, which severely limits the efficiency of local Monte Carlo simulations
in the presence of divergent spatial correlations. The archetypal, and spectacularly
successful, examples are found in the Ising and Potts models which, as first shown by
Fortuin and Kasteleyn~\cite{FK72a,FK72b,FK72c} and independently by Coniglio and
Klein~\cite{coniglio:80a}, admit a formulation in an extended space of spin and
auxiliary bond variables, now known as the Fortuin–Kasteleyn–Coniglio–Klein (FKCK)
representation. With the help of alternating updates in the spin and bond subspaces,
Swendsen and Wang~\cite{swendsen87} and later Wolff~\cite{wolff89} developed
rejection-free cluster Monte Carlo algorithms that achieve a fundamental acceleration
in the decorrelation of system configurations. The key to this success lies in the
coincidence of the onset of percolation of FKCK clusters with the thermal phase
transition of the spin model, as well as in the fact that the geometric properties of
the critical clusters mirror the critical correlations of the spin degrees of freedom
(for a recent review, see Ref.~\cite{coniglio:21}).

These algorithms work extremely well for the Potts model. Based on the
embedded-cluster trick proposed by Wolff~\cite{wolff89}, extensions to
continuous-spin models with analogous interactions are straightforward and similarly
effective. When moving beyond these paradigmatic examples, however, cluster
approaches often become more difficult to construct and/or do not work so
well. Although somewhat more general cluster-update frameworks have been
proposed~\cite{edwards:88a,kandel:90a,kandel:91a}, they do not yield efficient
algorithms in all situations, in particular for systems with frustrated
interactions. (Even greater challenges arise in the presence of additional disorder;
see Ref.~\cite{munster:23}.) An especially interesting case occurs for systems with
multi-spin interactions, where an extension of configuration space in terms of bond
variables is no longer particularly natural. Such models appear in several contexts,
including proposals for quantum computing architectures~\cite{pachos04,mizel04}, the
design of novel storage devices~\cite{savvidy15}, models exhibiting glassy dynamics
without quenched disorder~\cite{dimopoulos02}, and studies of metastable phases
following quenches~\cite{oike25}.

The simplest nontrivial example of such a system is the Baxter--Wu model, an Ising
model on the triangular lattice with three-spin interactions~\cite{wood72}. While, in
principle, a cluster algorithm based on freezing triangular plaquettes could be
devised following the general framework of Refs.~\cite{kandel:90a,kandel:91a}, the
only concrete proposal to date for the Baxter--Wu model is due to Novotny and
Evertz~\cite{novotny93}. Their approach effectively reduces the problem to the case
of pairwise interactions by freezing one of the three sublattices of the triangular
lattice and constructing clusters on the remaining two. Building upon this idea, Deng
\emph{et al.}~\cite{deng10} developed cluster-update schemes for generalized variants
of the problem, including versions with two distinct interaction strengths and others
incorporating three-spin interactions in a $q$-state Potts model. To date, however, a
detailed analysis of the approach of Novotny and Evertz has not been
presented. (Ref.~\cite{velonakis11} largely repeats the original construction using
different terminology, without offering significant further insight). It has also
been argued that the resulting clusters may not percolate precisely at the thermal
critical point~\cite{shchur10}, a phenomenon observed in several related
systems~\cite{saberi15,meyers09,picco09}. In other words, the percolation threshold
of the constructed clusters, $T_{\mathrm{p}}$, may differ from the actual critical
temperature, $T_{\mathrm{c}}$, implying that, asymptotically, the stochastic spin
clusters cannot serve to efficiently decorrelate configurations as the system size
increases.

Hence, a detailed percolation analysis of the original cluster construction
introduced in Ref.~\cite{novotny93} has so far been lacking in the literature. In the
present work, we close this gap in the understanding of cluster updates for spin
systems by performing a comprehensive study of the percolation properties of the
Novotny–Evertz clusters for the Baxter--Wu model. We find that these clusters indeed
percolate precisely at the thermal critical point and fully capture the thermal
critical behavior, thereby providing a firm justification for constructing cluster
algorithms based on this prescription. By investigating both multi-cluster and
single-cluster variants, we further demonstrate that these algorithms significantly
reduce autocorrelation times compared to single-spin flip updates, and we determine
the corresponding dynamical critical exponents.

The rest of this paper is organized as follows. In
Sec.~\ref{sec:Model_Method_Observables}, we introduce the Baxter--Wu model, describe
the Monte Carlo algorithms employed in our simulations, and define the observables
used in both the cluster and dynamical analyses. Section~\ref{sec:Results} contains
our main results. Specifically, Sec.~\ref{sec:Percolation} focuses on determining the
percolation threshold of the clusters and extracting the associated critical
exponents, while Sec.~\ref{sec:z} provides a detailed analysis of the dynamical
scaling behavior of the algorithm at equilibrium as a function of system
size. Finally, Sec.~\ref{sec:Conclusions} summarizes our main findings and outlines
potential directions for future work.

\section{Model and Numerics}
\label{sec:Model_Method_Observables}

\subsection{Model}
\label{sec:Model}

\begin{figure}[tb!]
  \centering
  \includegraphics[width=1.0\linewidth]{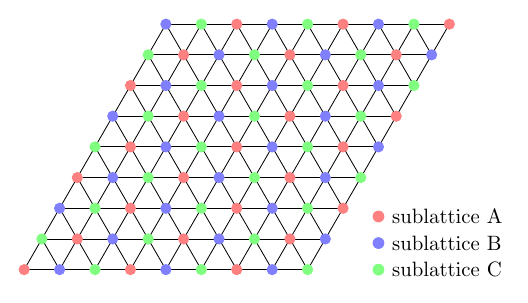}
  \caption{Example of a $9 \times 9$ triangular lattice with the three sublattices
    distinguished by color. Each spin interacts with its nearest neighbors on the
    other two sublattices, illustrating the triplet-interaction structure
    characteristic of the BW model.}
  \label{fig:lattice}
\end{figure}

We consider the spin-$1/2$ Baxter--Wu (BW) model, defined on a triangular lattice by
the Hamiltonian~\cite{wood72}
\begin{equation}\label{eq:H}
\mathcal{H} = -J\sum_{\langle ijk \rangle} \sigma_i \sigma_j \sigma_k,
\end{equation}
where the Ising spins $\sigma_i$ take values $\pm 1$, and the summation extends over
all elementary triangular plaquettes formed by nearest-neighbor triplets
$\langle ijk \rangle$. Throughout this work, we set the interaction strength $J=1$,
thereby fixing the energy scale, and we apply periodic boundary conditions. As
illustrated in Fig.~\ref{fig:lattice}, the triangular lattice decomposes naturally
into three interpenetrating sublattices (A, B, and C), such that each corner of a
triangle belongs to a different sublattice. For a system of linear size $L$, each
sublattice contains $N/3$ spins, where $N=L^2$ is the total number of lattice sites.

The characteristic three-spin interaction results in a loss of the overall
spin-inversion symmetry of the standard Ising model. On the other hand, a
simultaneous inversion of all spins on any two sublattices leaves the Hamiltonian
invariant, resulting in a fourfold degenerate ground state. Originally introduced by
Wood and Griffiths as a variation of the Ising model with three-spin interactions
that preserves self-duality~\cite{wood72}, the BW model was later solved exactly by
Baxter and Wu~\cite{baxter73,baxter74a,baxter74b,baxter_book}. It was shown to belong
to the same universality class as the four-state Potts model, exhibiting a continuous
phase transition with central charge $c=1$ (but without logarithmic corrections).

\subsection{Algorithms}
\label{sec:Algorithm}

For constructing stochastic clusters, we follow the original proposal of Novotny and
Evertz~\cite{novotny93}. Their approach starts from the observation that we know how
to construct cluster updates for standard Ising systems through the FKCK
representation. They hence suggested a transformation of the BW model into an
effective two-body problem achieved by freezing all spins on one of the three
sublattices (which is randomly selected at each update step). As a result, the spins
on the remaining two sublattices acquire effective two-body couplings given by
\begin{equation}
  J_{ij}^{\prime} = J(\sigma_{\perp (i,j), +1} + \sigma_{\perp (i,j), -1}),
\end{equation}
where $\sigma_{\perp(i,j),\pm1}$ denote the two spins on the frozen sublattice
opposite of the bond $(i,j)$.

\begin{figure}[tb!]
    \centering
    \includegraphics[width=1.0\linewidth]{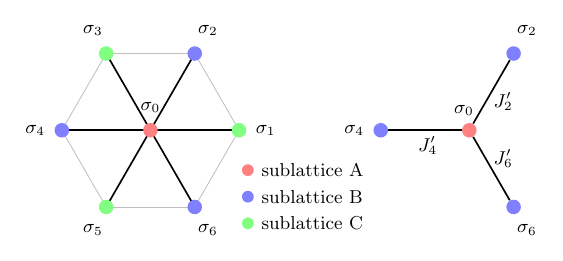}
    \caption{Local mapping of the triangular lattice onto a honeycomb lattice by
      freezing one of the three sublattices. In the example shown, the spins on the
      frozen C sublattice lead to effective pair interactions between the spins on
      the A and B sublattices, denoted by $J^{\prime}$. For the bond $(0,2)$ the
      opposite spins on sublattice C are $\sigma_{\perp(0,2),+1} = \sigma_3$ and
      $\sigma_{\perp(0,2),-1} = \sigma_1$.}
    \label{fig:cluster_modification}
\end{figure}

An example of this construction is illustrated in
Fig.~\ref{fig:cluster_modification}, where sublattice C is frozen, resulting in
renormalized couplings between the A spin $\sigma_0$ and the B spins $\sigma_2$,
$\sigma_4$, and $\sigma_6$. We note that the frozen sublattice and the resulting
honeycomb lattice are dual to each other. Since we set $J=1$ throughout, the
effective couplings $J_{ij}^\prime$ take values in $\{-2, 0, +2\}$, corresponding to
antiferromagnetic, diluted, and ferromagnetic bonds, respectively. Importantly, this
procedure introduces no frustration: the product of the couplings along any hexagon,
and hence around any closed loop, is non-negative.

As a consequence, the resulting effective Ising model with only pairwise interactions
on a diluted honeycomb lattice can be simulated using standard FKCK-based cluster
algorithms. The established proof of detailed balance for these algorithms for the
conventional Ising and Potts models therefore applies directly here. Moreover,
because the choice of the frozen sublattice is random, the dynamics are ergodic (the
probability of all clusters consisting of a single spin is nonzero) and hence must
converge to the correct equilibrium distribution according to the Markov
theorem~\cite{novotny93}.

If a spin $\sigma_i$ already belongs to a given cluster, a neighboring spin
$\sigma_j$ is eligible to be added to the cluster if the effective interaction is
satisfied, i.e., if $J_{ij}^{\prime} \sigma_i \sigma_j > 0$. Following the FKCK
rules, the probability of adding such a spin is given by
\[
  p_{\rm add} = 1 - \exp{(-2|J^{\prime}_{ij}|/T)},
\]
where $T$ is the system temperature. Bonds with $J_{ij}^\prime = 0$ can be retained
without explicitly considering a diluted lattice, as they will never be activated.  A
simplified multi-cluster implementation (in the spirit of the Swendsen–Wang
algorithm) proceeds as follows:
\begin{enumerate}[itemsep=1pt,parsep=1pt,leftmargin=17pt,label=(\arabic*)]
\item Randomly freeze one sublattice (e.g., sublattice C, as shown in
  Fig.~\ref{fig:cluster_modification}). Only spins on the remaining two sublattices
  are considered for the cluster construction.
\item Choose a seed spin that has not yet been assigned to a cluster.
\item For each of its three neighbors not already in a cluster, check whether the
  effective interaction condition $J_{ij}^{\prime} \sigma_i \sigma_j > 0$ is
  satisfied. If so, add the neighbor to the current cluster with probability
  $p_{\rm add} = 1 - \exp(-4/T)$.
\item Continue growing the cluster recursively by applying step 3 to newly added
  spins until no further spins can be added.
\item If any unassigned spins remain, return to step 2.
\item Once all $2N/3$ active spins have been assigned to clusters, flip each cluster
  with probability $1/2$. Return to step 1 for the next Monte Carlo step.
\end{enumerate}

A single-cluster update (analogous to the Wolff algorithm) can be implemented by
constructing a single cluster per Monte Carlo step, initiated from a randomly chosen
spin in one of the active sublattices. Each spin may participate in only one cluster
per step. Depending on the signs of the effective interactions, the resulting
clusters can be ferromagnetic or antiferromagnetic. However, due to the bipartite
nature of the honeycomb lattice, the fixed configuration of the frozen sublattice,
and the available spin values, clusters containing both ferromagnetic and
antiferromagnetic bonds are not possible.

\begin{figure}[bt!]
  \centering
  \includegraphics[width=1.0\linewidth]{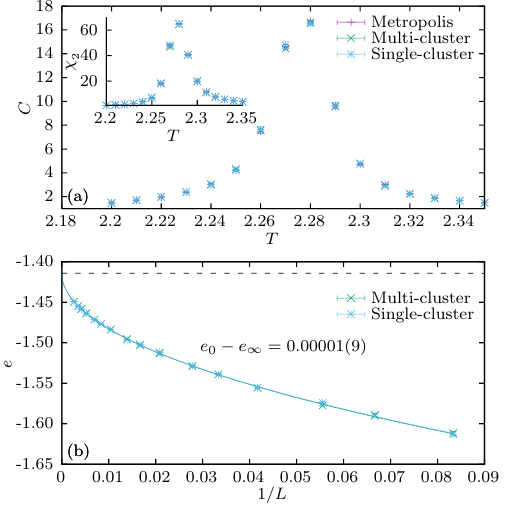}
  \caption{(a) Comparison of simulation results obtained using the Metropolis,
    single-cluster, and multi-cluster updates for the specific heat and magnetic
    susceptibility of the spin-$1/2$ BW model at various temperatures for a system of
    linear size $L=48$. The main panel shows the specific heat $C(T)$, while the
    inset displays the magnetic susceptibility $\chi_2(T)$. The excellent agreement
    between the methods confirms the correctness of the cluster-algorithm
    implementation.
    (b) Finite-size scaling of the energy at the critical point. The exact result $e_0=-\sqrt{2}$~\cite{baxter73,baxter74a,baxter74b} is indicated by the dashed line.}
  \label{fig:comparison_spin1half}
\end{figure}

In a single Monte Carlo step, $2N/3$ spins---those not on the frozen sublattice---are
considered for updates, thereby defining one Monte Carlo time step for the
multi-cluster update. It is clear from the construction of the system that for the
sublattice decomposition, and hence the validity of the presented algorithms, $L$ and
$N = L^2$ must be multiples of three. The correctness of the cluster-update scheme and our implementation was ascertained, among other checks, by comparing various quantities between the Metropolis and cluster simulations. The result of such comparisons are summarized in 
Fig.~\ref{fig:comparison_spin1half}. In particular, panel (a) shows a the specific heat
$C$ [main panel, Eq.~\eqref{eq:c}] and the magnetic susceptibility $\chi_2$ [inset,
Eq.~\eqref{eq:chi}] obtained using the multi- and single-cluster algorithms as well as
results from the standard Metropolis algorithm~\cite{metropolis,landau_book} for a
system of linear size $L = 48$. All data are fully compatible within statistical errors. Figure~\ref{fig:comparison_spin1half}(b) illustrates the finite-size scaling behavior of the energy and its convergence towards the exact asymptotic value $e_0 = -\sqrt{2}$~\cite{baxter73,baxter74a,baxter74b} for the two cluster algorithms. Since the multi-cluster and single-cluster simiuations are statistically independent, we performed a joint fit to the two data sets using the ansatz~\cite{amit_book}
\[
    e(L) = e_\infty + b\,L^{-(d - 1/\nu)},
\]
where $e_\infty$ denotes the thermodynamic-limit value of the energy, $b$ is a fitting parameter, $d = 2$ is the spatial dimensionality, and $\nu = 2/3$ corresponds to the Baxter-Wu universality class.
Allowing both $e_\infty$ and $d - 1/\nu$ to vary, we obtain $e_0 - e_\infty = 0.0002(4)$ and $d - 1/\nu = 0.502(4)$. Fixing $d - 1/\nu = 0.5$ and refitting yields $e_0 - e_\infty = 0.00001(9)$. Finally, fixing $e_\infty = -\sqrt{2}$ and fitting only the remaining parameters gives $d - 1/\nu = 0.5000(8)$. All results are in excellent agreement with the expected values.

In our production runs, we performed $10^5$ Monte Carlo steps for sampling at the
smallest system size, $L = 12$, with an additional $10^4$ steps used for
equilibration. For larger systems, the number of Monte Carlo steps was scaled
proportionally to $N/12^2$, roughly accounting for the expected dynamical critical
exponent $z \approx 2$ for local algorithms. Specifically, we considered systems with
linear sizes $12 \le L \le 384$. For each size, $20$ independent realizations were
simulated, and statistical analysis was performed using the jackknife
method~\cite{efron92}. Critical behavior was extracted via least-squares fitting,
with a lower cut-off $L \ge L_{\rm min}$ chosen to account for scaling
corrections. Fit quality was assessed using the standard $\chi^2$
test~\cite{press92}, and fits were deemed acceptable if the goodness-of-fit parameter
satisfied $Q \ge 10\%$~\cite{press92}.

\subsection{Observables}
\label{sec:Observables}

While the discussion above demonstrates that the proposed cluster update is formally
correct, it is not \emph{a priori} clear whether it is effective in alleviating
critical slowing down. One necessary condition is that the clusters just begin to
percolate at the point of the thermal phase transition; otherwise, they will
asymptotically include either very few or nearly all spins, effectively reducing the
update to a local move. To investigate the percolation properties of the BW clusters
introduced above, we studied standard observables from percolation
theory~\cite{stauffer_introduction_1994}, namely the wrapping probability
$P_{\rm wrap}$, the average cluster size $S$, and the percolation strength
$P_{\infty}$.

The wrapping (or spanning) probability $P_{\rm wrap}$ is defined as the probability
that at least one cluster spans the periodic boundaries of the system, wrapping
around the lattice and reconnecting with itself. In the thermodynamic limit, the
wrapping probability $P_{\rm wrap}$ becomes a discontinuous function of temperature,
taking the value $P_{\rm wrap} = 0$ above the percolation transition temperature
$T_{\rm p}$ and $P_{\rm wrap} = 1$ below it. This discontinuity signals the
appearance of percolating clusters for $T < T_{\rm p}$. In contrast, for finite
systems, $P_{\rm wrap}(T)$ is a smooth, continuous function. Nonetheless, curves
corresponding to different system sizes are expected to intersect at a common
point---modulo finite-size effects---marking the percolation transition. Depending on
the spatial direction in which clusters wrap, various definitions of $P_{\rm wrap}$
can be employed~\cite{newman00,martins03,akritidis23}. In this study, a cluster is
considered to percolate if it wraps around and reconnects to itself in either the
horizontal or vertical direction, or in both directions. As a dimensionless quantity,
$P_{\rm wrap}$ is expected to obey the standard finite-size scaling (FSS)
form~\cite{stauffer_introduction_1994}:
\begin{equation}\label{eq:pwrap_fss}
  P_{\rm wrap} (t,L) = \Tilde{P}_{\rm wrap}( t L^{1/ \nu}),
\end{equation}
where $\Tilde{P}_{\rm wrap}$ is a universal scaling function, $t = (T-T_{\text{p}})/T_{\text{p}}$ is the reduced temperature, and $\nu$ is the
critical exponent associated with the divergence of the correlation length.

The average cluster size is defined as
\begin{equation}\label{eq:S}
S = \frac{\sum_s n_s s^2}{\sum_s n_s s},
\end{equation}
where $n_s$ denotes the number of clusters of size
$s$~\cite{stauffer_introduction_1994}. In the thermodynamic limit, excluding the
(infinite) percolating cluster from the sums in Eq.~\eqref{eq:S} causes $S$ to peak
near the percolation transition. This intermediate maximum arises because at high
temperatures the system consists predominantly of small clusters, whereas below the
percolation threshold $T_{\rm p}$ most spins belong to the macroscopic percolating
cluster (which is excluded from the sums). To reproduce this behavior in finite
systems, the largest cluster is excluded from each measurement. As a result, $S$
develops a peak at a pseudocritical temperature, which approaches $T_{\rm p}$ as the
system size increases. In the critical regime, $S$ obeys the FSS
form:
\begin{equation}\label{eq:S_fss}
  S(t,L) = L^{\gamma / \nu} \Tilde{S}(t L^{1/ \nu}), 
\end{equation}
where $\gamma/\nu$ is the ratio of critical exponents associated with the divergence
of the average cluster size, analogous to the finite-size scaling exponent of the magnetic susceptibility~\cite{stauffer_introduction_1994}.  In previous numerical studies it was
noted that this exclusion of percolating clusters (or an analogous subtraction in the
susceptibility) introduces significant scaling corrections in Ising-like
systems~\cite{janke05,akritidis23}. In the present work, we therefore use a
definition of $S$ that includes all clusters, without omitting the largest or
spanning clusters. The downside of this approach, of course, is that $S$ no longer
exhibits a maximum, and must instead be evaluated either at the fixed temperature
$T_{\rm p}$ or at a separately determined sequence of pseudocritical points.

The percolation strength $P_{\infty}$ denotes the probability that a randomly
selected spin belongs to the percolating cluster. It is computed as the fraction of
sites comprising the largest, system-spanning cluster. In the thermodynamic limit,
$P_{\infty} = 0$ above the percolation threshold, indicating the absence of a
spanning cluster, while below $T_{\rm p}$ it grows continuously, approaching
$P_{\infty} = 1$ as $T \to 0$, where all spins belong to a single connected
cluster. Serving as an order parameter---analogous to the magnetization in a thermal
phase transition---$P_{\infty}$ captures the onset of long-range connectivity. Its
FSS form is given by~\cite{stauffer_introduction_1994}
\begin{equation}\label{eq:P_fss}
  P_{\infty}(t,L) = L^{-\beta / \nu} \Tilde{P}_{\infty}(t L^{1/ \nu}),
\end{equation}
	where $\beta/\nu$ is the finite-size scaling exponent associated with the order parameter.

In order to study the thermodynamic properties of the system and, by extension, the
dynamical properties of the algorithm, we measured the internal energy $E$, from
which the specific heat is obtained via the standard fluctuation–dissipation
relation as
\begin{equation}
\label{eq:c}
C = (\langle E^2 \rangle - \langle E \rangle^2)/(NT^2).
\end{equation}
Magnetic ordering was characterized by evaluating the magnetization on each of the
three sublattices, denoted by $M_{\rm A}$, $M_{\rm B}$, and $M_{\rm C}$. From these,
we define two commonly used order parameters:
\begin{align}
    \label{eq:op1} m_1 &= (|m_{\rm A}| + |m_{\rm B}| + |m_{\rm C}|)/3,\\
    \label{eq:op2} m_2 &= \sqrt{(m_{\rm  A}^2+m_{\rm B}^2+m_{\rm C}^2)/3},
\end{align}
where $m_x = M_x/(N/3)$ for each sublattice $x = {\rm A}$, ${\rm B}$, and
${\rm C}$~\cite{novotny81,costa04b,liu21}. In the thermodynamic limit, both $m_1$ and
$m_2$ approach unity in the fully ordered ground states and vanish in the completely
disordered (paramagnetic) state.  The associated magnetic susceptibilities are
defined as
\begin{equation}\label{eq:chi}
\chi_i = \frac{N \left(\langle m_i^2 \rangle - \langle m_i \rangle^2\right)}{T}, \quad i = 1, 2,
\end{equation}
quantifying the fluctuations in the respective order parameters.

\begin{figure}[tb!]
    \centering
    \includegraphics[width=1.0\linewidth]{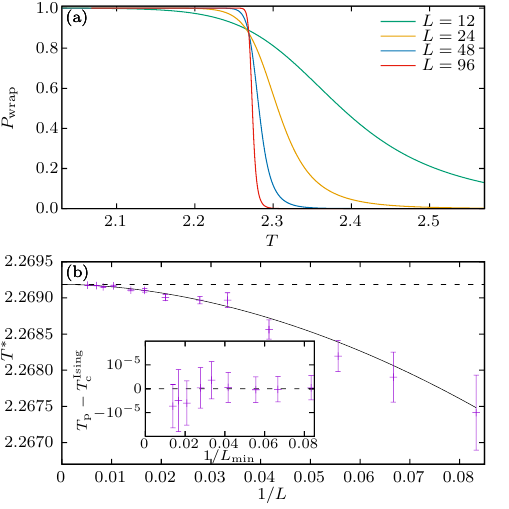}
    \caption{(a) Wrapping probability as a function of temperature for different
      system sizes. (b) Extrapolation of the percolation temperature $T_{\rm p}$ from
      the crossings of the wrapping probability curves according to
      Eq.~\eqref{eq:scaling_crossing}. The horizontal dashed line indicates the exact
      critical temperature $T_{\rm c}$ of the BW model. The inset shows the
      difference $T_{\rm p} - T_{\rm c}$ as a function of the minimum system size
      $L_{\rm min}$ included in the fits. The excellent agreement between the two
      temperatures supports identifying the percolation transition with the thermal
      phase transition.}
    \label{fig:Pwrap_crossing_spin1half}
\end{figure}

As shown by Wood and Griffiths~\cite{wood72}, the BW model is self-dual and
consequently exhibits the same critical temperature as the regular square-lattice
Ising model~\cite{merlini72,baxter_book},
\begin{equation}
\label{eq:Tc}
T_{\rm c}^{\rm Ising} = \frac{2}{\log{(\sqrt{2}+1)}} \approx 2.269\, 185\, 314 \cdots.
\end{equation}
All simulations were therefore conducted directly at this critical
temperature. Furthermore, the use of the single-histogram reweighting
technique~\cite{ferrenberg:88a} enabled efficient extraction of observables over a
range of temperatures near criticality. Our independent estimate of the percolation
temperature, reported below, additionally justifies this choice of simulation
temperature \emph{a posteriori}, confirming the self-consistency of the analysis.

\section{Results}
\label{sec:Results}

\subsection{Percolation analysis}
\label{sec:Percolation}

To determine the percolation temperature, we analyzed the wrapping probabilities for
pairs of system sizes $(L, 2L)$. As shown in
Fig.~\ref{fig:Pwrap_crossing_spin1half}(a), the wrapping probabilities approximately
cross at a common point. According to standard arguments of FSS, these crossing
points are expected to follow the law~\cite{amit_book}
\begin{equation}\label{eq:scaling_crossing}
    T^\ast(L) = T_{\rm p} + c_pL^{-(\omega+1/\nu)},
\end{equation}
where $T^\ast(L)$ denotes the crossing temperature, $T_{\rm p}$ is the percolation
temperature in the thermodynamic limit $L\to\infty$, $c_p$ is a non-universal
amplitude, and $\omega$ is the correction-to-scaling exponent~\cite{alcaraz97,
  alcaraz99}.  Figure~\ref{fig:Pwrap_crossing_spin1half}(b) shows $T^\ast$ as a
function of $1/L$, while the inset displays the difference
$T_{\rm p} - T_{\rm c}^{\rm Ising}$ as a function of the minimum system size
$L_{\rm min}$ included in the fits.  From this analysis, we obtain the final estimate
\begin{equation}
\label{eq:T_p}
T_{\rm p} = 2.269\, 186(5),
\end{equation}
corresponding to $T_{\rm p} - T_{\rm c}^{\rm Ising} = (0 \pm 5) \times 10^{-6}$,
where we used $L_{\rm min} = 12$ and the resulting exponent estimate is
$\omega+1/\nu = 1.9(1)$. (We will come back to this estimate further below in the
present section.)  Within numerical uncertainty, these results demonstrate that the
percolation temperature of the clusters coincides with the critical temperature of
the model.

\begin{figure}[tb!]
    \centering
    \includegraphics[width=1.0\linewidth]{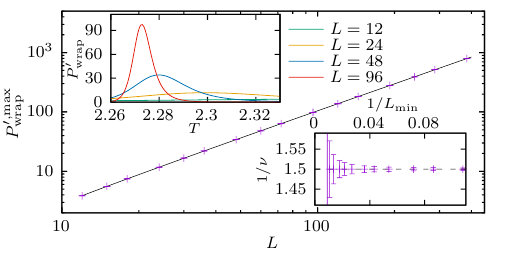}
    \caption{FSS of the derivatives of the wrapping probability
      evaluated at their respective maxima (main panel). The inset in the top left
      shows the temperature dependence of the derivative of the wrapping probability
      for several system sizes. The bottom-right inset displays the estimated
      exponent $1/\nu$ as a function of $1/L_{\rm min}$ for fits including
      corrections and a free $\omega$ parameter; the horizontal dashed line marks the
      exact value $1/\nu = 3/2$.} \label{fig:Pwrap_derivative_scaling_spin1half}
\end{figure}

The derivative of the wrapping probability is also of interest, as it provides an
independent route to estimating the critical exponent $\nu$. It was computed using a
three-point numerical differentiation scheme based on histogram
reweightinge~\cite{ferrenberg:88a} with a temperature step of $10^{-6}$. The peak
value of this derivative is expected to follow the FSS
relation~\cite{stauffer_introduction_1994}
\begin{equation}\label{eq:scaling_Pwrap_derivative}
  P^{\prime,\mathrm{max}}_{\rm wrap} (L) = c_{P^\prime}L^{1/\nu}\left(1+d_{P^\prime} L^{-\omega}\right).
\end{equation}
where $c_{P^\prime}$ and $d_{P^\prime}$ are non-universal fitting constants, and the
superscript ``max'' indicates evaluation at the maximum of $P'_{\rm wrap}$.
Figure~\ref{fig:Pwrap_derivative_scaling_spin1half} shows representative fits
according to Eq.~\eqref{eq:scaling_Pwrap_derivative}.  Three types of fits were
performed: (i) without corrections, excluding small system sizes to mitigate
finite-size effects; (ii) including corrections with a fixed
$\omega = 2$~\cite{alcaraz97, alcaraz99}; and (iii) allowing $\omega$ to vary freely.
For the first case, excluding the correction terms, the estimates for $1/\nu$ are
several standard deviations away from the expected value of $3/2$; for example, when
$L_{\rm min} = 120$, $1/\nu=1.507(4)$. Only for $L_{\rm min} \geq 192$ do we get
values consistent with 3/2: for $L_{\rm min} = 192$ specifically, $1/\nu = 1.504(6)$.
From the fits with a correction-to-scaling exponent set to $\omega=2$, fits become
acceptable for $L_{\rm min}\geq 36$ and we again see a discrepancy from the exact
value. Starting from $L_{\rm min}\geq 120$, we find values in agreement with the
expected one: for example, with $L_{\rm min}=120$, we find
$1/\nu=1.500(4)$. Interestingly, for all values of $L_{\rm min}$, corrections seem to
remain relevant, since $d_{P^\prime}$ does not disappear within errors, which would
be expected as one eliminates smaller systems.  With a free correction exponent,
$1/\nu = 1.500(2)$ and $\omega = 1.0(2)$, for $L_{\rm min}=12$. Importantly, despite
the uncertainty in corrections, the estimates of $1/\nu$ remain consistent across all
fitting procedures.

In view of the above results, one may wonder about the consistency of values of the
correction-to-scaling exponent $\omega$ obtained from
Eqs.~\eqref{eq:scaling_crossing} and~\eqref{eq:scaling_Pwrap_derivative}.  From the
fit to the crossing temperatures [Eq.~\eqref{eq:scaling_crossing}], we find
$\omega + 1/\nu = 1.9(1)$. Using any of our estimates of $1/\nu$, this corresponds to
$\omega \approx 0.4$. However, previous studies by Alcaraz \emph{et
  al.}~\cite{alcaraz97, alcaraz99} report $\omega = 2$. On the other hand, the above
quoted fit for $ P^{\prime,\mathrm{max}}_{\rm wrap} (L)$ yields $\omega = 1.0(2)$,
while we do not find evidence in favor of $\omega = 2$ in our data.  These
discrepancies suggest that the present data set may not be sufficient to reliably
resolve subleading corrections to scaling, indicating that additional correction
terms are likely relevant but cannot be captured within our accessible range of
system sizes. Consequently, in the remainder of the analysis, we focus on fits
including the correction exponent as a free parameter. Only when the constants in
front of the correction term $L^{-\omega}$ become minuscule do we consider additional
fitting processes.

\begin{figure}[tb!]
    \centering
    \includegraphics[width=1.0\linewidth]{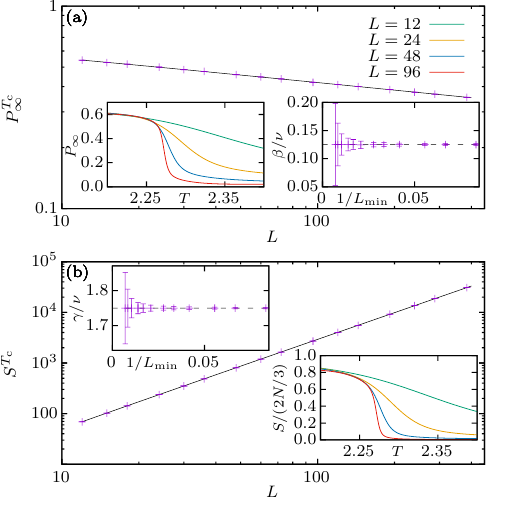}
    \caption{FSS of (a) the percolation strength $P_{\infty}$ and (b)
      the average cluster size $S$, including the largest cluster. Both observables
      are evaluated at the Ising-model critical temperature~\eqref{eq:Tc}. Power-law
      fits yield the critical exponent ratios $\beta/\nu$ and $\gamma/\nu$,
      respectively. The insets show the estimated exponents as a function of
      $L_{\rm min}$, with horizontal dashed lines indicating the exact values
      $\beta/\nu = 1/8$ and $\gamma/\nu = 7/4$. Also included are insets with the
      corresponding curves of $P_\infty$ and the normalized cluster size $S/(2N/3)$.}
    \label{fig:Sstar_Pinfinity_scaling_spin1half}
\end{figure}

Finally, by analyzing the scaling behavior of the percolation strength $P_\infty$ and
the average cluster size $S$, both evaluated at the critical temperature $T_{\rm c}$
of Eq.~(\ref{eq:Tc})---which according to our results equals the percolation
temperature
$T_\mathrm{p}$---, we obtain estimates for the critical exponent ratios
$\beta/\nu$ and
$\gamma/\nu$. These quantities are expected to follow the FSS
relations~\cite{stauffer_introduction_1994}
\begin{align}
    \label{eq:Pinfinity_scaling} P_\infty^{T_\mathrm{c}}(L) &= c_{P}L^{-\beta/\nu}\left(1+d_{P} L^{-\omega}\right),\\
    \label{eq:Sast_scaling} S^{T_\mathrm{c}}(L) &= c_{S}L^{\gamma/\nu}\left(1+d_{S} L^{-\omega}\right),
\end{align}
where $c_{P}$, $c_{S}$ and $d_{P}$, $d_{S}$ are non-universal fitting constants.  The
corresponding fits are shown in Fig.~\ref{fig:Sstar_Pinfinity_scaling_spin1half},
panels (a) and (b) for $P_\infty^{T_\mathrm{c}}$ and $S^{T_\mathrm{c}}$, respectively.  As discussed in
Sec.~\ref{sec:Observables}, our definition of $S$ includes all clusters;
consequently, this quantity does not exhibit a peak as a function of temperature.
From the fits to Eqs.~\eqref{eq:Pinfinity_scaling} and~\eqref{eq:Sast_scaling}, we
obtain $\beta/\nu = 0.1251(9)$ and $\gamma/\nu = 1.750(2)$, with $L_{\rm min} = 12$,
both in excellent agreement with the exact values $\beta/\nu = 1/8$ and
$\gamma/\nu = 7/4$ characteristic of the BW universality class. The corresponding
corrections-to-scaling exponents are $\omega = 1.3(8)$ and $1.0(3)$. For the
$\beta/\nu$ fits, the correction amplitude $d_P$ is consistently zero within
errors. Fits performed without corrections yield compatible results for
$L_{\rm min} \geq 120$ only; for example,
$\beta/\nu = 0.1244(6)$, $0.1243(7)$, $0.124(1)$, $0.123(1)$ for
$L_{\rm min} = 120$, $144$, $192$, $240$, respectively, all consistent with the expected
value $1/8$.

\subsection{Dynamical critical exponent}
\label{sec:z}

To estimate the dynamical critical exponent $z$ of the cluster algorithms, we first
compute the integrated autocorrelation times $\tau$ for three observables: the energy
[Eq.~\eqref{eq:H}] and the two order parameters defined in Eqs.~\eqref{eq:op1}
and~\eqref{eq:op2}. The corresponding autocorrelation times, denoted as $\tau_e$,
$\tau_{m_1}$, and $\tau_{m_2}$, are evaluated at the critical temperature. Their
scaling behavior with system size is expected to follow the FSS
ansatz~\cite{barkema_book}
\begin{equation}\label{eq:tau_scaling}
\tau_x (L) = c_xL^{z_x}\left(1+d_x L^{-\omega}\right),
\end{equation}
where $c_x$ and $d_x$ are non-universal fitting coefficients, and
$z_x \equiv z_x^{\rm int}$ denotes the (integrated) dynamical critical exponent
associated with observable $x \in \{e, m_1, m_2\}$. We note that the
observable-independent exponent $z^{\rm exp}$, which characterizes the scaling of
exponential autocorrelation times, is not considered in the present analysis.

\begin{figure}[tb!]
    \centering
    \includegraphics[width=1.0\linewidth]{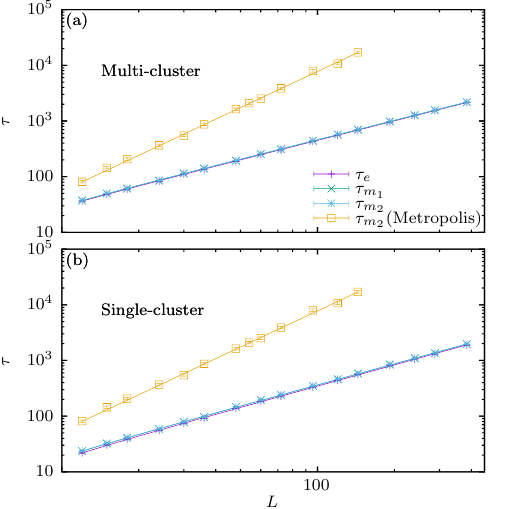}
    \caption{Scaling behavior of the integrated autocorrelation times for the energy
      and the two order parameters defined in Eqs.~\eqref{eq:op1} and
      \eqref{eq:op2}. The dynamical critical exponent $z$ is extracted via fits to
      the scaling form in Eq.~\eqref{eq:tau_scaling}. Panel (a) shows results for the
      multi-cluster update, while panel (b) corresponds to the single-cluster
      update. For comparison, Metropolis results for $m_2$ are also included. All
      curves exhibit an overall linear trend, with the single-cluster data displaying
      a slightly steeper slope than the multi-cluster results.}
    \label{fig:z_cutoff_spin1half}
\end{figure}

The integrated autocorrelation times $\tau_x$ are defined via a summation of the
normalized autocorrelation functions,
\begin{equation}
\label{eq:norm_autocorr_fun}
A_x(t^\prime) = \frac{\langle x(t) x(t+t^\prime) \rangle - \langle x(t) \rangle \langle x(t+t^\prime) \rangle }{\langle x(t)^2 \rangle - \langle x(t) \rangle^2},
\end{equation}
where $x$ denotes the observable under consideration. In practice, the natural
estimator $\hat{A}_x(t^\prime)$ of the autocorrelation function is computed directly
from the time series of measurements. The integrated autocorrelation time is then
estimated employing a summation cut-off~\cite{madras88, sokal97, janke08}:
\begin{equation}\label{eq:tau_cut-off}
\hat{\tau}_x = I_x(k_{\rm max}^{(x)}) = \frac{1}{2} + \sum_{t^\prime=1}^{k^{(x)}_{\rm max}} \hat{A}_x(t^\prime).
\end{equation}
The cut-off $k_{\rm max}^{(x)}$ is determined self-consistently as the smallest lag
$t^\prime$ satisfying $k_{\rm max}^{(x)} > 6 \hat{\tau}_x$~\cite{madras88}. This
criterion provides a useful tradeoff between the systematic error for too small
cut-off $k_{\rm max}^{(x)}$ and a divergent variance of the estimator for
$k_{\rm max}^{(x)}\to\infty$. For the single-cluster algorithm, the definition of one
Monte Carlo time step must take into account that only a single cluster is
constructed per update. To ensure comparable time units with the multi-cluster
algorithm, we scale time by a factor $\langle C\rangle / (2N/3)$, corresponding to
the average fraction of the $2N/3$ active sites at each step that is updated in one
single-cluster step with average cluster size $\langle C\rangle$. This normalization
allows for a direct comparison of autocorrelation times between the two update
schemes.

Our main results for the autocorrelation times of the cluster updates are presented
in Fig.~\ref{fig:z_cutoff_spin1half}, which also shows, for comparison, the
autocorrelation behavior of the $m_2$ observable calculated with the Metropolis
algorithm. The dynamical critical exponent $z$ is extracted by fitting the
system-size dependence of the integrated autocorrelation times $\tau_x$ to the FSS
form given in Eq.~\eqref{eq:tau_scaling}. The resulting estimates for the single- and
multi-cluster algorithms are summarized in Table~\ref{tab:z_estimates}. For fits with
a free corrections-to-scaling exponent $\omega$, the terms $d_x$ consistently vanish
for all $x$, for both the single- and the multi-cluster algorithm. However, when
$\omega$ is fixed to 2~\cite{alcaraz97,alcaraz99}, we find a non-zero amplitude if
using small $L_{\rm min}$. In these cases including corrections, the results stated
in Table~\ref{tab:z_estimates} are for $L_{\rm min}=12$.
For the case without corrections, $L_{\rm min}$ is 24 for the multi-cluster and 48
for the single-cluster case. Similar to the results in Sec.~\ref{sec:Percolation},
our estimates of $z$ are consistent regardless of the method used and across all
observables.  From the fits without corrections to scaling, we obtain average
estimates of $z = 1.162(3)$ for the multi-cluster update
[Fig.~\ref{fig:z_cutoff_spin1half}(a)] and $z = 1.251(5)$ for the single-cluster
update [Fig.~\ref{fig:z_cutoff_spin1half}(b)]. The latter lies at the lower end of
the estimate reported by Novotny and Evertz, $z = 1.37(10)$~\cite{novotny93}, which
was obtained using a single-cluster implementation and based on the root-mean-square
sublattice magnetization [see Eq.~\eqref{eq:op2}]. Both cluster algorithms clearly
outperform the Metropolis method, for which we find $z = 2.16(3)$, demonstrating the
substantial reduction of critical slowing down. The Metropolis result is, within
statistical uncertainty, consistent with that of the Ising
model~\cite{nightingale98,liu23,bisson25}.

\begin{table}[tb!]
  \caption{Final estimates of the dynamical critical exponent $z$ of the integrated
    autocorrelation times for the multi- and single-cluster algorithms (labeled MC
    and SC, respectively) in the BW model, based on the three observables analyzed in
    this work. We show results obtained with fixed corrections using $\omega=2$, with
    free corrections with variable $\omega§$, and without scaling corrections. [see
    Eq.~\eqref{eq:tau_scaling}]. For comparison, the last two rows include the
    estimate from Ref.~\cite{novotny93} and our results obtained using the Metropolis
    algorithm.}
\begin{ruledtabular}
    \begin{tabular}{clccc}
Algorithm           & \multicolumn{1}{c}{Fit type} & $z_e$     & $z_{m_1}$ & $z_{m_2}$ \\ \hline
\multirow{3}{*}{MC} & free corr.                   & 1.157(15) & 1.152(15) & 1.152(15) \\
                    & fixed corr.                  & 1.163(3)  & 1.158(3)  & 1.159(3)  \\
                    & no corr.                     & 1.166(3)  & 1.161(3)  & 1.162(3)  \\ \hline
\multirow{3}{*}{SC} & free corr.                   & 1.254(22) & 1.246(22) & 1.246(22) \\
                    & fixed corr.                  & 1.255(6)  & 1.247(6)  & 1.247(6)  \\
                    & no corr.                     & 1.257(5)  & 1.248(5)  & 1.248(5)  \\ \hline
Ref.~\cite{novotny93}       & no corr.                     & -         & -         & 1.37(10)  \\ \hline
Metropolis          & no corr.                     & -         & -         & 2.16(3)  
    \end{tabular}\label{tab:z_estimates}
\end{ruledtabular}
\end{table}

\section{Conclusions}
\label{sec:Conclusions}

We have investigated the percolation properties of clusters in the spin-$1/2$
Baxter–Wu model using the construction scheme proposed by Novotny and
Evertz~\cite{novotny93}, in which one sublattice is frozen and clusters are built on
the remaining two. Implementing a multi-cluster update within this framework, we
first verified that, within numerical accuracy, the clusters percolate precisely at
the known critical temperature of the model. Furthermore, we extracted the critical
exponent ratios $1/\nu$, $\beta/\nu$, and $\gamma/\nu$ associated with the cluster
observables and found them to be consistent with the known thermal values of the
universality class. These findings confirm the validity of the algorithm and its
suitability for simulating the Baxter–Wu model.

To assess the efficiency of the cluster algorithm, we analyzed its dynamical behavior
by estimating the critical exponent $z$ from the scaling of integrated
autocorrelation times for various observables, considering both multi- and
single-cluster implementations. The integrated autocorrelation time was computed
using a self-consistent cut-off method. This analysis yields estimates of
$z = 1.251(5)$ for the single-cluster variant and $z = 1.162(3)$ for the
multi-cluster algorithm. These values lie close to the lower bound imposed by the
Li–Sokal inequality, $z \ge \alpha / \nu = 2/\nu - d = 1$~\cite{li:89}, indicating
that these cluster algorithms operate near optimal efficiency for the Baxter–Wu
model.

Our present study lays the groundwork for a similar percolation and dynamical scaling
analysis in the spin-$1$ generalization of the Baxter–Wu model, which includes a
chemical potential~\cite{dias17,jorge20,vasilopoulos22,macedo23}. In this setting,
the presence of zero spins may hinder the effectiveness of cluster algorithms,
particularly at larger crystal-field strengths $\Delta$, where the density of zero
spins increases significantly. Interestingly, this model allows for clusters that
combine both ferromagnetic and antiferromagnetic interactions. Since zero spins
cannot be easily included in the cluster construction, a pure cluster algorithm is no
longer ergodic and must be complemented by local single-spin updates in a hybrid
scheme---a strategy that has proven successful for other spin-$1$
models~\cite{zierenberg17}. This generalized model is also believed to host a
pentacritical point~\cite{dias17,jorge20,vasilopoulos22,macedo23}, leading to a phase
diagram containing both first- and second-order transition lines. This raises several
intriguing questions: (i) Do the clusters still percolate at the transition points?
(ii) How does the dynamical critical exponent $z$ behave near the putative
multicritical point? (iii) Does $z$ depend systematically on $\Delta$?

Another promising avenue is the study of the gonihedric
model~\cite{johnston96,espriu97}, where the presence of plaquette interactions
complicates cluster construction. In this case, a modified cluster
approach---splitting the lattice into frozen and active sublattices to induce
effective interactions---may provide a viable strategy for future investigations.

\begin{acknowledgments}
  We would like to thank Wolfhard Janke for several fruitful discussions on this subject. The numerical calculations reported in this paper were performed at the
  High-Performance Computing cluster CERES of the University of Essex. The work of A.V. and N.G.F. was supported by the Engineering and Physical Sciences Research
  Council (Grant No. EP/X026116/1).
\end{acknowledgments}

\bibliography{bib.bib}

\end{document}